\documentclass[a4paper, oneside, twocolumn, notitlepage, 10pt]{extarticle_ecoc}
\usepackage{ecoc}

\usepackage[acronym]{glossaries}
\newacronym{acf}{ACF}{autocorrelation function}
\newacronym{air}{AIR}{achievable information rate}
\newacronym{app}{APP}{a posteriori probability}
\newacronym{ase}{ASE}{amplified spontaneous emission}
\newacronym{ask}{ASK}{amplitude-shift keying}
\newacronym{awgn}{AWGN}{additive white Gaussian noise}

\newacronym{b2b}{B2B}{back-to-back}
\newacronym{baa}{BAA}{Blahut-Arimoto Algorithm}
\newacronym{bpcu}{bpcu}{bits per channel use}

\newacronym{cd}{CD}{chromatic dispersion}
\newacronym{cdc}{CDC}{chromatic dispersion compensation}
\newacronym{cdf}{cdf}{cumulative density function}
\newacronym{clt}{CLT}{central limit theorem}
\newacronym{coi}{COI}{channel of interest}
\newacronym{cpan}{CPAN}{correlated phase and additive noise}
\newacronym{cspr}{CSPR}{carrier to signal power ratio}
\newacronym{cscg}{CSCG}{circularly symmetric complex Gaussian}
\newacronym{csi}{CSI}{channel state information}
\newacronym{cwt}{CWT}{continouus-wave tone}

\newacronym{dbp}{DBP}{digital backpropagation}
\newacronym{dsp}{DSP}{digital signal processing}
\newacronym{dtft}{DTFT}{discrete time Fourier transform}

\newacronym{dd}{DD}{direct detection}
\newacronym{dft}{DFT}{discrete Fourier transform}
\newacronym{dmd}{DMD}{differential mode delay}
\newacronym{dmgd}{DMGD}{differential mode group delay}
\newacronym{dgd}{DGD}{differential group delay}
\newacronym{dt}{DT}{decision tree}

\newacronym{eepn}{EEPN}{equalization-enhanced phase noise}
\newacronym{ep}{EP}{expectation propagation}
\newacronym{epi}{EPI}{entropy power inequality}
\newacronym{erp}{ERP}{enhanced regular perturbation}
\newacronym{evd}{EVD}{Eigenvalue decomposition}

\newacronym{fer}{FER}{frame error rate}
\newacronym{ffe}{FFE}{feedforward equalizer}
\newacronym{fft}{FFT}{fast Fourier transform}
\newacronym{fmf}{FMF}{few-mode fiber}
\newacronym{fwhm}{FWHM}{full-width half-maximum}
\newacronym{fwm}{FWM}{four-wave mixing}

\newacronym{ghq}{GHQ}{Gauss-Hermite-Quadrature}
\newacronym{gmi}{GMI}{generalized mutual information}
\newacronym{gvamp}{GVAMP}{generalized \gls{vamp}}

\newacronym{ici}{ICI}{intercarrier interference}
\newacronym{idr}{IDR}{ideal data remodulation}
\newacronym{idra}{IDRA}{ideally distributed Raman amplification}
\newacronym{iid}{i.i.d.}{independent and identically distributed}
\newacronym{im}{IM}{intensity modulation}
\newacronym{isi}{ISI}{intersymbol interference}
\newacronym{iss}{ISS}{intersymbol sample}

\newacronym{jd}{JD}{joint detection}
\newacronym{jdd}{JDD}{joint detection and decoding}
\newacronym{jdabaa}{DABAA}{dynamic-assignment Blahut-Arimoto Algorithm}

\newacronym{kk}{KK}{Kramers-Kronig}

\newacronym{lhs}{l.h.s.}{left-hand side}
\newacronym{lmmse}{LMMSE}{linear minimum-mean-square-error}
\newacronym{ls}{LS}{least squares}
\newacronym{lstm}{LSTM}{long short-term memory}
\newacronym{lp}{LP}{linearly polarized}
\newacronym{lut}{LUT}{look-up table}

\newacronym{map}{MAP}{maximum a-posteriori}
\newacronym{mcf}{MCF}{multi-core fiber}
\newacronym{md}{MD}{modal dispersion}
\newacronym{mgdm}{MGDM}{mode-group-division multiplexing}
\newacronym{mdg}{MDG}{mode-dependent gain}
\newacronym{mdl}{MDL}{mode-dependent loss}
\newacronym{mdm}{MDM}{mode-division multiplexing}
\newacronym{me}{ME}{Manakov equation}
\newacronym{mi}{MI}{mutual information}
\newacronym{mir}{MIR}{mutual information rate}
\newacronym{mimo}{MIMO}{multiple-in multiple-out}
\newacronym{miso}{MISO}{multiple-in single-out}
\newacronym{ml}{ML}{machine learning}
\newacronym{mmf}{MMF}{multi-mode fiber}
\newacronym{mmse}{MMSE}{minimum-mean-squared-error}
\newacronym{mpc}{MPC}{minimum-phase criterion}
\newacronym{mse}{MSE}{mean-squared error}
\newacronym{mzm}{MZM}{Mach-Zehnder modulator}

\newacronym{nlse}{NLSE}{nonlinear Schrödinger equation}
\newacronym{nmse}{NMSE}{normalized MSE}
\newacronym{nn}{NN}{neural network}

\newacronym{ofdm}{OFDM}{orthogonal frequency-division multiplexing}

\newacronym{pam}{PAM}{pulse-amplitude modulation}
\newacronym{pdf}{pdf}{probability density function}
\newacronym{pn}{PN}{phase noise}
\newacronym{pnc}{PNC}{phase noise compensation}
\newacronym{pll}{PLL}{phase-locked loop}
\newacronym{pmf}{pmf}{probability mass function}
\newacronym{psd}{PSD}{power spectral density}
\newacronym{psk}{PSK}{phase-shift keying}

\newacronym{qam}{QAM}{quadrature-amplitude modulation}

\newacronym{roadm}{ROADM}{reconfigurable optical add-drop multiplexer}
\newacronym{rp}{RP}{regular perturbation}
\newacronym{rrc}{RRC}{root-raised cosine}
\newacronym{rv}{RV}{random variable}

\newacronym{sd}{SD}{seperate detection}
\newacronym{sdd}{SDD}{separate detection and decoding}
\newacronym{sdm}{SDM}{space-division multiplexing}
\newacronym{ssfm}{SSFM}{split-step Fourier method}
\newacronym{simo}{SIMO}{single-in multiple-out}
\newacronym{sinr}{SINR}{signal to interference and noise ratio}
\newacronym{siso}{SISO}{single-in single-out}
\newacronym{sld}{SLD}{square-law detector}
\newacronym{snr}{SNR}{signal-to-noise ratio}
\newacronym{smf}{SMF}{single-mode fiber}
\newacronym{sonr}{SONR}{signal and offset to noise ratio}
\newacronym{spa}{SPA}{sum-product algorithm}
\newacronym{spm}{SPM}{self-phase modulation}
\newacronym{ssb}{SSB}{single side-band}
\newacronym{ssbi}{SSBI}{signal-signal beat interference}
\newacronym{ssmf}{SSMF}{standard single-mode fiber}
\newacronym{svd}{SVD}{singular value decomposition}
\newacronym{svm}{SVM}{state vector machine}

\newacronym{te}{TE}{transverse electric}
\newacronym{tm}{TM}{transverse magnetic}
\newacronym{tp}{TP}{transparent propagation}

\newacronym{urr}{URR}{unidistant Rayleigh ring}

\newacronym{vamp}{VAMP}{vector approximate message passing}

\newacronym{wdm}{WDM}{wavelength-division multiplexing}
\newacronym{wlg}{w.l.g.}{without loss of generality}

\newacronym{xpm}{XPM}{cross-phase modulation}

\newacronym{zf}{ZF}{zero forcing}

\usepackage{amssymb} 
\usepackage{bm}

\usepackage{subfigure}
\usepackage{enumitem}
\usepackage[normalem]{ulem}

\usepackage{tikz}
\usetikzlibrary{positioning}
\usetikzlibrary{calc}
\usetikzlibrary{arrows}
\usepackage{pgfplots}

\usepackage{algorithm}
\usepackage{algpseudocode}

\usepackage[detect-all]{siunitx}
\DeclareSIUnit\bpcu{bpcu}

\usepackage{algorithm}
\usepackage{algpseudocode}

\definecolor{matlabBlue}{rgb}{0.00000,0.44700,0.74100}%
\definecolor{matlabRed}{rgb}{0.85000,0.32500,0.09800}%
\definecolor{matlabYellow}{rgb}{0.92900,0.69400,0.12500}%
\definecolor{matlabPurple}{rgb}{0.49400,0.18400,0.55600}%
\definecolor{matlabGreen}{rgb}{0.46600,0.67400,0.18800}%
\definecolor{matlabLightBlue}{rgb}{0.30100,0.74500,0.93300}%
\definecolor{matlabMagenta}{rgb}{0.63500,0.07800,0.18400}%

\definecolor{matlab1}{rgb}{0.00000,0.44700,0.74100}%
\definecolor{matlab2}{rgb}{0.85000,0.32500,0.09800}%
\definecolor{matlab3}{rgb}{0.92900,0.69400,0.12500}%
\definecolor{matlab4}{rgb}{0.49400,0.18400,0.55600}%
\definecolor{matlab5}{rgb}{0.46600,0.67400,0.18800}%
\definecolor{matlab6}{rgb}{0.30100,0.74500,0.93300}%
\definecolor{matlab7}{rgb}{0.63500,0.07800,0.18400}%

\definecolor{mycolor1}{rgb}{0.00000,0.44700,0.74100}%
\definecolor{mycolor2}{rgb}{0.85000,0.32500,0.09800}%
\definecolor{mycolor3}{rgb}{0.92900,0.69400,0.12500}%
\definecolor{mycolor4}{rgb}{0.49400,0.18400,0.55600}%
\definecolor{mycolor5}{rgb}{0.46600,0.67400,0.18800}%
\definecolor{mycolor6}{rgb}{0.30100,0.74500,0.93300}%
\definecolor{mycolor7}{rgb}{0.63500,0.07800,0.18400}%

\newcommand{\ratiopilots}{{\ensuremath{\rho}}}

\renewcommand{\j}{{\ensuremath{\mathrm{j}}}}
\newcommand{\e}{{\ensuremath{\mathrm{e}}}}
\renewcommand{\vec}[1]{{\ensuremath{\bm{#1}}}}
\newcommand{\mat}[1]{{\ensuremath{\bm{#1}}}}
\newcommand{\transpose}{{\ensuremath{\mathsf{T}}}}
\newcommand{\hermitian}{{\ensuremath{\mathsf{H}}}}
\renewcommand{\v}[1]{\vec{#1}}
\newcommand{\m}[1]{\mat{#1}}

\newcommand{\figref}[1]{Fig.~\ref{#1}}

\tikzset{
    fnode/.style={
        draw, fill=black, minimum size = 10pt
        },
    vnode/.style={
        draw, shape = circle, minimum size = 25pt, inner sep = 0pt
        },
    msgarrow/.style={
        dashed, color = gray!60!black
        },
}

\newcommand{\commentalex}[1]{}
\newcommand{\commentgerhard}[1]{}

\DeclareSIUnit\gigabaud{GBaud}
\DeclareSIUnit\bpcu{bpcu}

\pgfplotsset{compat=1.18}
\newcommand\figscale{1} %

\begin{document}
\selectlanguage{english}    %

\title{Feedforward and Iterative Phase Noise Compensation \\ for Channels with Chromatic Dispersion}%

\author{Alex Jäger, Gerhard Kramer}

\maketitle
\begin{strip}
    \begin{author_descr}
    
        Institute for Communications Engineering, Technical University of Munich,
        \textcolor{blue}{\uline{alex.jaeger@tum.de}}
    \end{author_descr}
\end{strip}

\let\origfootnotemark\footnotemark
\renewcommand\footnotemark{}
\renewcommand\footnoterule{}

\begin{strip}
    \begin{ecoc_abstract}
        Equalization-enhanced phase noise is avoided by applying phase noise compensation (PNC) before chromatic dispersion compensation. Feedforward and iterative PNC algorithms based on expectation propagation are proposed. Both achieve information rates close to channels without phase noise for 100 GBaud 64-QAM and 10,000 km of fiber. \hfill\textcopyright2026 The Author(s)
    \end{ecoc_abstract}
\end{strip}

\let\footnotemark\origfootnotemark
\section{Introduction}
A \gls{cdc} module changes \gls{pn} statistics, resulting in \gls{eepn} \cite{Shieh:08}. Subsequent \gls{pnc} can partially mitigate \gls{eepn}, but the distortions tend to increase with laser linewidth, symbol rate, and fiber length \cite{Farhoudi:12,Arnould:19}. In contrast, the \gls{pn} before \gls{cdc} is independent of distance and decreases with symbol rate. 

We introduce two algorithms for \gls{pnc} before \gls{cdc}. First, a feedforward receiver avoids excessive latency, and, second, an iterative, \gls{ep}-based receiver improves performance. Both substantially outperform \gls{cdc} before \gls{pnc} for the \gls{snr} and \gls{pn} variances that we studied. 

\stepcounter{footnote}\footnotetext{
The model \eqref{eq:pn-model} has one receiver sample per transmit symbol. However, many \gls{pn} models have large bandwidth, which means that multiple samples per symbol can be useful \cite{Ghozlan:13,Ghozlan:17}. We remark that, if one wishes to differentiate between transmitter and receiver \gls{pn}, then one requires at least two pilot tones at conjugate frequency pairs \cite{Colavolpe:11,You:24,Zhang:24,Sekizuka:25}.}
\section{System Model}
Consider transmit symbols $\{M_i\}_{i=1}^n$ that are uniformly and independently drawn from a finite constellation, e.g., \gls{qam}. The constellation has mean zero and variance $\mathrm{E}[|M_i|^2]=\sigma_m^2$. Consider the $n$-dimensional vectors $\v M=[M_1,\dots,M_n]^\transpose$ and $\v 1=[1,\dots,1]^\transpose$, where the latter represents pilot symbols. We transmit $\v X=\v M+ \rho \v 1$ for some real, positive $\rho$, i.e., we use a pilot tone. This gives larger \glspl{air} in our simulations than interleaved pilots do. The transmit power is normalized to $\sigma_m^2+\rho^2=1$, i.e., the pilot offset $\rho$ reduces $\sigma_m^2$.

The output of a \gls{cd} channel is then
\begin{equation}
    \v Z = \m H \v X = \m H (\v M + \rho \v 1)
\end{equation}
with unitary \gls{cd} matrix $\m H$. We assume that $\v Z$ experiences the \gls{pn}$^{\thefootnote}$
\begin{equation}
    Y_i = \e^{\j\Theta_i}Z_i+N_{i}
    \label{eq:pn-model}
\end{equation}
where $\v N$ is \gls{awgn} with variance $\sigma_n^2$.
We model the string $\{\Theta_i\}_{i=1}^n$ as Wiener \gls{pn} where the increments $\Delta_i=\Theta_i-\Theta_{i-1}$ are independent Gaussian with mean zero and variance $\sigma_\theta^2$ and $\Theta_1\sim\mathcal{U}(-\pi,\pi)$. The ratio $1/\sigma_n^2$ is referred to as the \gls{snr}.

\section{Receiver Structures}

\noindent\emph{\Gls{cdc} before \gls{pnc}:}
The output of a receiver that performs \gls{cdc} first can be formulated as 
\begin{equation}
    \v Y' = \m H^\hermitian \v Y ,\qquad \v Y'' = f(\v Y')
\end{equation}
where $f(\cdot )$ is a \gls{pnc} algorithm, e.g., the Viterbi-Viterbi algorithm \cite{Viterbi:83} or blind phase search \cite{Pfau:09}. Consider the surrogate model
\begin{equation}
    Y'_i \approx \e^{\j\Theta'_i}X_i+N'_i .
\end{equation}
If the receiver knows $\tilde{\v Y}=\v Y'-\v N'$, it may compute an estimate $\hat{\Theta}_i'=\angle\left(\tilde{Y}_iX_i^*\right)$ and output $Y''_i = \e^{-\j\hat{\Theta}_i'}Y'_i$. This approach is similar to \gls{idr} proposed in \cite{Arnould:19}.

After $f(\cdot )$, the receiver models the remaining distortions as \gls{awgn} $\v N''$ with variance $\sigma_{n''}^2$, i.e., $\v Y'' \approx \v X+\v N''$, which is equivalent to using a Gaussian surrogate decoding metric 
\begin{align*}
    q(\v y|\v x)
    = \mathcal{N}_{\mathbb{C}}(\v y'';\v x,\sigma_{n''}^2\m I)
    = \prod_{i=1}^n \frac{e^{-|y_i''-x_i|^2/\sigma_{n''}^2}}{\pi \sigma_{n''}^2}.
\end{align*}
This approach causes \gls{eepn} \cite{Shieh:08,Jung:25,Geiger:25}.
\ \\

\begin{figure}[t]
    \centering
    \scalebox{0.9}{
    \begin{tikzpicture}
        \node (x) at (0,0) {$\v X$};
        \node[draw, right = .5cm of x] (cd) {CD};
        \node[draw, right = .5cm of cd] (pn) {PN};
        \node[draw, right = 1cm of pn] (pnc) {PNC};
        \node[draw, right = .5cm of pnc] (cdc) {CDC};
        \node[right = 1cm of cdc] (yx) {$\v Y_x$};
        \node[right = .5cm of cdc,fill=black,inner sep = 0,shape=circle,minimum width = 3] (dot) {};
        \node[draw, dashed, below = .5cm of dot] (den) {Demapper};
        \node[draw, dashed, below = 1cm of cdc] (cd2) {CD};
        
        \draw[->] (x)--(cd);
        \draw[->] (cd)--node[midway,above] {$\v Z$} (pn);
        \draw[->] (pn)--node[midway, above] {$\v Y$} (pnc);
        \draw[->] (pnc)--node[midway,above] {$\v Y_z$} (cdc);
        \draw[->] (cdc)--(yx);
        \draw[->, dashed] (dot)--(den);
        \draw[->, dashed] (den) |- node[midway,right] {$\hat{\v X}$} (cd2);
        \draw[->, dashed] (cd2)  -| node[midway,left] {$\hat{\v Z}$} (pnc);
    \end{tikzpicture}
    }
    \caption{System model with a feedforward receiver (solid) and iterative (dashed) receiver.}
    \label{fig:systemmodel}
\end{figure}
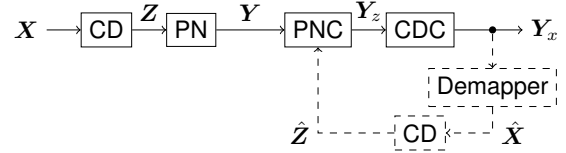

\noindent\emph{Feedforward \gls{pnc}:}
\figref{fig:systemmodel} shows a system that performs \gls{pnc} first. The goal is to compute a ``good'' vector $\v Y_z$ for the \gls{cdc} module, i.e., with the statistics
\begin{equation}
    \v Y_z\approx \v Z+\v N_z
\end{equation}
where $\v N_z$ is \gls{awgn} with small variance $\sigma_{n_z}^2$. The \Gls{cdc} computes 
\begin{equation}
    \label{equ:y_x}
    \v Y_x = \m H^\hermitian \v Y_z \approx \v X+\v N_x
\end{equation}
where $\v N_x$ has the same statistics as $\v N_z$.

We design the \gls{pnc} by using a surrogate $q(\v z)$ motivated by the \gls{clt}, i.e., the density $q(\v z)$ is a product of offset-\gls{cscg} densities with means $\hat{\v z}=\m H \, \rho \v 1=\rho\v 1$ and variance $\sigma_m^2$ (note that CD is an all-pass filter and transfers a constant signal $\rho \v 1$ unchanged).
\ \\

\noindent\emph{Iterative Phase Noise and ISI Compensation:}
We also study an iterative algorithm based on \gls{ep}\cite{minka2001family} to update $q(\v z)$ by using the \emph{non-Gaussian} constellation prior $P(\v x)$, see the dashed elements in \figref{fig:systemmodel}. A demapper uses $\v Y_x$ and $P(\v x)$ to update $\hat{\v x}$ so that 
\begin{equation}
    \v X\approx \hat{\v x}+\v W
\end{equation}
where $\v W$ is \gls{awgn} that represents uncertainty about the estimate. The updated surrogate $q(\v x)$ is thus offset-\gls{cscg} with mean $\hat{\v x}$ and variance $\sigma_w^2$. The filtered $\hat{\v z}=\m H\hat{\v x}$ is the mean of the updated prior $q(\v z)$, which is passed to the \gls{pnc}. One now iterates. This approach is similar to turbo equalization \cite{Douillard:95:iterative}, but there is no decoder in the turbo loop.

\section{PN Compensation}
The \gls{pnc} uses the channel output $\v Y=\v y$ and mismatched prior $q(\v z)$. Consider the mismatched posterior
\begin{equation}
    r(\v z|\v y) = \frac{q(\v z)p(\v y|\v z)}{\int_{\mathbb{C}^n} q(\v z')p(\v y|\v z') \mathrm{d}\v z'}
\end{equation}
where
\begin{align}
    p(\v y|\v z) &= \int\nolimits_{[-\pi,\pi)^n} p(\v \theta) p(\v y|\v \theta, \v z)\mathrm{d}\v \theta\\
    p(\v y|\v\theta,\v z) &=  \prod\nolimits_{i=1}^n\mathcal{N}_\mathbb{C}\left(y_i;z_i\e^{\j\theta_i},\sigma_n^2\right).
\end{align}
Observe that $r(\v z|\v y)$ is the true posterior if the true prior was $q(\v z)$. We use moment matching \cite{minka2001family} to compute an offset-\gls{cscg} approximation $q(\v z|\v y)=\mathcal{N}_\mathbb{C}(\v z;\v \mu_1,\sigma_1^2\m I)$ of $r(\v z|\v y)$ with
\begin{equation}
    \v \mu_1 = \mathrm{E}\left[\v Z\right],\qquad 
    \sigma_1^2 = \frac{1}{n}\sum_{i=1}^n\mathrm{E}\left[|Z_i-\mu_{1,i}|^2\right]
\end{equation}
where the expectations are with respect to $r(\v z|\v y)$. Note that $\v \mu_1$ is the \gls{mmse} estimate with posterior $r(\v z|\v y)$.

To compute means, we use
\begin{equation}
    \mu_{1,i} = \int\nolimits_{\mathbb{C}^n} z_i\, r(\v z|\v y)\mathrm{d}\v z
\end{equation}
and likewise for the variance. We use the \gls{spa} and approximate integrals with a von Mises-based algorithm, which has linear complexity in $n$ \cite{Jaeger:26,Colavolpe:05,Szczecinski:20}.

Similar to Bayes' rule, we compute $q(\v y|\v z)\propto q(\v z|\v y)/q(\v z)$. A surrogate $q(\v y)$ of $p(\v y)$ is a normalization constant, and we absorb it into a $\int q(\v y|\v z)\mathrm{d}\v z=1$ normalization. As $q(\v z|\v y)$ and $q(\v z)$ are offset-\gls{cscg}, $q(\v y|\v z)$ is an offset-\gls{cscg} with mean and variance
\begin{equation}
    \v y_z = \frac{\sigma_w^2\v \mu_1-\sigma_1^2\hat{\v z}}{\sigma_w^2-\sigma_1^2}\qquad\sigma_{n_z}^2=\frac{\sigma_w^2\sigma_1^2}{\sigma_w^2-\sigma_1^2}.
\end{equation}
This step removes the information that $q(\v z)$ contributed to $q(\v z|\v y)$, such that $q(\v y|\v z)$ only provides new evidence on $\v z$. It is hence called "extrinsic" in \gls{ep} \cite{minka2001family}. Note that the extrinsic variance can become negative, but we did not see this happening in our simulations.

\section{Demapper}
The demapper uses the prior $P(\v x)$, estimated mean $\v y_x$, and estimated variance $\sigma_{n_x}^2=\sigma_{n_z}^2$ of the offset-\gls{cscg} $q(\v y|\v x)$. The variance is preserved by \gls{cdc}, because $\m H^\hermitian$ is unitary. The demapper computes an offset-\gls{cscg} approximation $q(\v x|\v y)=\mathcal{N}_\mathbb{C}(\v x;\v \mu_2,\sigma_2^2\m I)$ of
\begin{equation}
    r(\v x|\v y) = \frac{q(\v y|\v x)P(\v x)}{\sum_{\v x'}q(\v y|\v x')P(\v x')}    
\end{equation}
by moment matching. Similar to \gls{pnc}, we compute the extrinsic $q(\v x)\propto q(\v x|\v y)/q(\v y|\v x)$, which is an offset-\gls{cscg} with mean and variance
\begin{equation}
    \hat{\v x} = \frac{\sigma_{n_x}^2\v \mu_2-\sigma_2^2\v y_x}{\sigma_{n_x}^2-\sigma_2^2}\qquad\sigma_w^2=\frac{\sigma_{n_x}^2\sigma_2^2}{\sigma_{n_x}^2-\sigma_2^2}.
\end{equation}

\section{Information Rates}
The \gls{gmi}
\begin{equation}
    I_\mathrm{gmi} = \mathrm{E}\left[\log_2\frac{q(\v Y|\v M)}{\sum\limits_{\v m'} P(\v m')q(\v Y|\v m')}\right] \leq I(\v M;\v Y)
\end{equation}
is an \gls{air} \cite{Scarlett:20}. The decoding metric $q(\v y|\v m)$ is obtained from $q(\v y|\v x)$ by subtracting $\v y_m=\v y_x-\rho\v 1$.

\section{Simulations}

\begin{table}[t]
    \centering
    \scalebox{0.9}{
    \begin{tabular}{c|c}
         Symbol rate & \SI{100}{\gigabaud}\\
         \hline
         \gls{cd} $\beta_2$ & \SI{-21.7}{\pico\second\squared\per\kilo\meter}\\
         \hline
         Constellation & 64-QAM \\
         \hline
         \# symbols/sequence & $2^{16}$\\
         \hline
         \# sequences & $64$\\
         \hline
         Fiber length L& \SI{10000}{\kilo\meter}\\
         \hline
         \gls{snr} & \SI{13}{\dB}\\
         \hline
         $\sigma_\theta^2$ & $\left\{10^{-5},10^{-4}\right\}$
    \end{tabular}
    }
    \caption{Simulation parameters unless specified otherwise. \gls{pn} variances correspond to almost \SI{160}{\kilo\hertz} and \SI{1.6}{\mega\hertz} linewidth, respectively.}
    \label{tab:simparams}
\end{table}

We simulate transmission over a channel with \gls{cd} and \gls{pn} using the parameters in Tab. \ref{tab:simparams}.

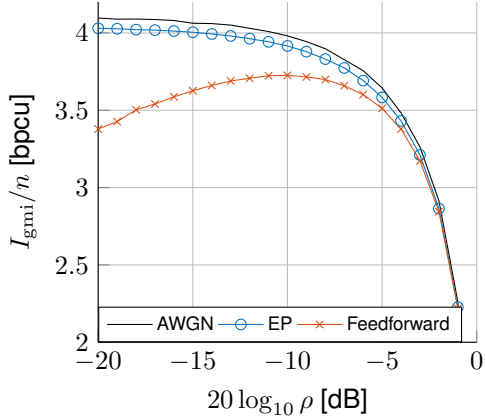
\begin{figure}[t]
    \centering
    \scalebox{\figscale}{
    \begin{tikzpicture}
    \begin{axis}[%
    width=5cm,
    height=4.5cm,
    at={(0,0)},
    scale only axis,
    xmin=-20,
    xmax=0,
    xlabel={$20\log_{10}\ratiopilots$ [dB]},
    ymin=2,
    ymax=4.2,
    ylabel={$I_\mathrm{gmi}/n$ [bpcu]},
    axis x line*=bottom,
    axis y line*=left,
    xmajorgrids,
    ymajorgrids,
    legend style={at={(0,0)},anchor=south west,nodes={scale=0.7, transform shape}},
    legend columns = 3,
    ]           
        \addplot [color=black]
          table[]{figures/data/13dB_awgn_air_v_k_uniform_offset_64.tsv};
        \addlegendentry{AWGN};
        
        \addplot [color=mycolor1, mark = o]
          table[]{figures/data/13dB_1e-4_air_v_k_cd_uniform_superposed_64_EP.tsv};
        \addlegendentry{EP};

        \addplot [color=mycolor2, mark = x]
          table[]{figures/data/13dB_1e-4_air_v_k_cd_uniform_superposed_64.tsv};
        \addlegendentry{Feedforward};

    \end{axis}
\end{tikzpicture}%
    }
    \caption{\gls{gmi} vs. offset for $\sigma_\theta^2=10^{-4}$.}
    \label{fig:air_v_rho}
\end{figure}

\noindent\emph{AIR vs. Offset Power:} \figref{fig:air_v_rho} compares the \gls{gmi} of the feedforward and \gls{ep} receivers to the \gls{awgn} channel with $\sigma_\theta^2=10^{-4}$, 64-\gls{qam}, and \SI{13}{\dB} \gls{snr}. The feedforward receiver is best at $\rho\approx\SI{-10}{\dB}$, but its maximum rate is relatively far from the \gls{awgn} curve. The \gls{ep} receiver required between 5 and 7 iterations for small $\rho$, and fewer as $\rho$ increases. \gls{ep} operates close to the \gls{awgn} curve, despite the large laser linewidth.

\begin{figure}[t]
    \centering
    \scalebox{\figscale}{
    \begin{tikzpicture}
    \begin{axis}[%
    width=5cm,
    height=4.5cm,
    at={(0,0)},
    scale only axis,
    xmin=-5,
    xmax=25,
    xlabel={\gls{snr} [dB]},
    ymin=0,
    ymax=7,
    ylabel={$I_\mathrm{gmi}/n$ [bpcu]},
    axis x line*=bottom,
    axis y line*=left,
    xmajorgrids,
    ymajorgrids,
    xtick={-5,0,5,10,15,20,25},
    ytick={0,1,2,3,4,5,6},
    legend style={at={(0,1)},anchor=north west,nodes={scale=0.7, transform shape}},
    legend columns = 2,
    ]           
        
    \addplot [color=mycolor1, mark = o] %
        table[]{figures/data/air_v_snr_ep_64_qam_1e-5.tsv};
    \addlegendentry{EP, $10^{-5}$};

    \addplot [color=mycolor1, mark = x]
        table[]{figures/data/air_v_snr_ep_64_qam_1e-4.tsv};
    \addlegendentry{EP, $10^{-4}$};
        
    \addplot [color=mycolor2, mark = ot]
        table[]{figures/data/air_v_snr_ff_64_qam_1e-5.tsv};
    \addlegendentry{FF, $10^{-5}$};
        
    \addplot [color=mycolor2, mark = x]
        table[]{figures/data/air_v_snr_ff_64_qam_1e-4.tsv};
    \addlegendentry{FF, $10^{-4}$};

    \addplot [color=mycolor3, mark=o]
        table[]{figures/data/air_v_snr_idr_64_qam_1e-5.tsv};
    \addlegendentry{IDR, $10^{-5}$};

    \addplot [color=mycolor3, mark = x]
        table[]{figures/data/air_v_snr_idr_64_qam_1e-4.tsv};
    \addlegendentry{IDR, $10^{-4}$};

    \addplot [color=black]
        table[]{figures/data/air_v_snr_awgn_64_qam.tsv};
    \addlegendentry{AWGN};

    \end{axis}
\end{tikzpicture}%
    }
    \caption{\gls{gmi} vs. \gls{snr}.}
    \label{fig:air_v_snr}
\end{figure}
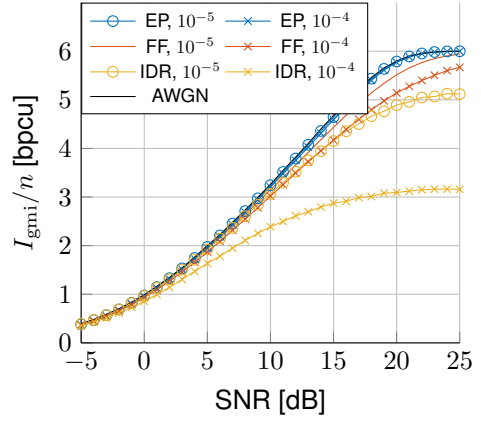

\noindent\emph{AIR vs. SNR:}
We next optimize $\rho$ for each \gls{snr}.
\figref{fig:air_v_snr} compares to the \gls{mi} of an \gls{awgn} channel with 64-\gls{qam} and $\rho=0$.

\Gls{idr} with \gls{eepn} exhibits large loss, saturating near \SI{5}{\bpcu} and \SI{3}{\bpcu}. In contrast, the feedforward receiver that reverses the \gls{pnc} and \gls{cdc} order operates close to the \gls{awgn} curve. The \gls{ep} rates almost coincide with the \gls{awgn} channel rates, but require many iterations for small \gls{snr}. At most 9 iterations are needed at intermediate and high \gls{snr} (larger than \SI{1}{\dB}).

\begin{figure}[t]
    \centering
    \vspace{0.4cm}
    \scalebox{\figscale}{
    \begin{tikzpicture}
    \begin{axis}[%
    width=5cm,
    height=4.5cm,
    at={(0,0)},
    scale only axis,
    xmin=1,
    xmax=15,
    xlabel={Fiber length [\SI{1000}{km}]},
    xtick = {1,5,10,15},
    ymin=2,
    ymax=4.5,
    ylabel={$I_\mathrm{gmi}/n$ [bpcu]},
    axis x line*=bottom,
    axis y line*=left,
    xmajorgrids,
    ymajorgrids,
    ytick={1,1.5,2,2.5,3,3.5,4},
    legend style={at={(0,0)},anchor=south west,nodes={scale=0.7, transform shape}},
    legend columns = 2,
    ]

    \addplot [color=mycolor1, mark = o]
            table[]{figures/data/air_v_L_ep_64_qam_snr_13dB_1e-5.tsv};
        \addlegendentry{EP, $10^{-5}$};
        
    \addplot [color=mycolor1, mark = x]
            table[]{figures/data/air_v_L_ep_64_qam_snr_13dB_1e-4.tsv};
        \addlegendentry{EP, $10^{-4}$};
        
    \addplot [color=mycolor2, mark = o]
            table[]{figures/data/air_v_L_ff_64_qam_snr_13dB_1e-5.tsv};
        \addlegendentry{FF, $10^{-5}$};
        
    \addplot [color=mycolor2, mark = x]
            table[]{figures/data/air_v_L_ff_64_qam_snr_13dB_1e-4.tsv};
        \addlegendentry{FF, $10^{-4}$};

    \addplot [color=mycolor3, mark = o]
            table[]{figures/data/air_v_L_idr_64_qam_snr_13dB_1e-5.tsv};
        \addlegendentry{IDR, $10^{-5}$};
        
    \addplot [color=mycolor3, mark = x]
            table[]{figures/data/air_v_L_idr_64_qam_snr_13dB_1e-4.tsv};
        \addlegendentry{IDR, $10^{-4}$};
        
    \end{axis}
\end{tikzpicture}%
    }
    \caption{GMI vs. fiber length.}
    \label{fig:air_v_L}
\end{figure}
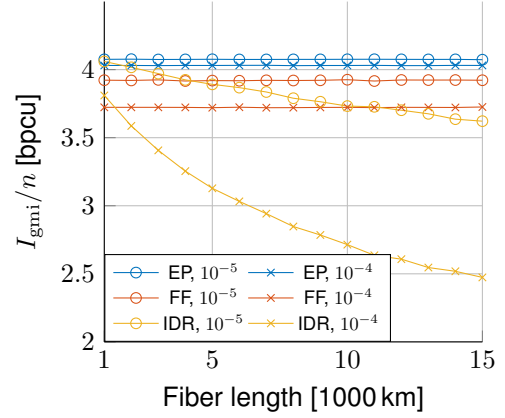

\figref{fig:air_v_L} plots the \gls{gmi} against the fiber length. \gls{idr} exhibits a logarithmic decay of \gls{gmi}; this is expected because \gls{eepn} causes an \gls{snr} penalty proportional to the fiber length \cite{Arnould:19}. Remarkably, the feedforward and \gls{ep} receiver performances remain almost constant vs. $L$. This suggests that the post-\gls{pnc} distortions are close to \gls{iid} \gls{cscg}, so the \gls{cdc} does not enhance or color them. 

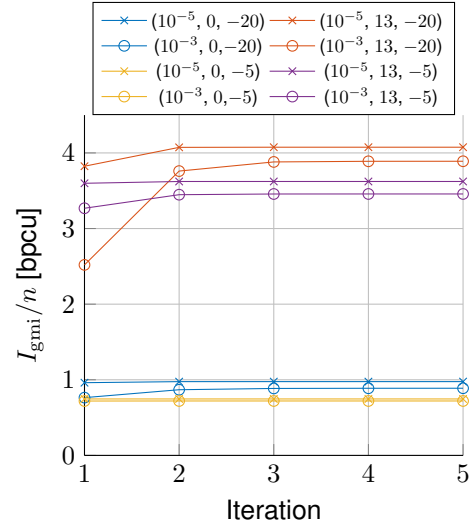
\begin{figure}[t!]
    \centering
    \vspace{0.4cm}
    \scalebox{\figscale}{
    \begin{tikzpicture}
    \begin{axis}[%
    width=5cm,
    height=4.5cm,
    at={(0,0)},
    scale only axis,
    xmin=1,
    xmax=5,
    xlabel={Iteration},
    ymin=0,
    ymax=4.5,
    ylabel={$I_\mathrm{gmi}/n$ [bpcu]},
    axis x line*=bottom,
    axis y line*=left,
    xmajorgrids,
    ymajorgrids,
    ytick={0,1,2,3,4},
    legend style={at={(0.5,1)},anchor=south,nodes={scale=0.7, transform shape}},
    legend columns = 2,
    ]           
        
        \addplot [color=mycolor1, mark = x] %
            table[]{figures/data/air_v_iter_64_qam_1e-5_10000km_0dB_-20dB.tsv};
        \addlegendentry{($10^{-5}$, $0$, $-20$)};
    
        \addplot [color=mycolor2, mark = x]
            table[]{figures/data/air_v_iter_64_qam_1e-5_10000km_13dB_-20dB.tsv};
        \addlegendentry{($10^{-5}$, $13$, $-20$)};
            
        \addplot [color=mycolor1, mark = o]
            table[]{figures/data/air_v_iter_64_qam_1e-3_10000km_0dB_-20dB.tsv};
        \addlegendentry{($10^{-3}$, $0$,$-20$)};
            
        \addplot [color=mycolor2, mark = o]
            table[]{figures/data/air_v_iter_64_qam_1e-3_10000km_13dB_-20dB.tsv};
        \addlegendentry{($10^{-3}$, $13$, $-20$)};
    
        \addplot [color=mycolor3, mark = x] %
            table[]{figures/data/air_v_iter_64_qam_1e-5_10000km_0dB_-5dB.tsv};
        \addlegendentry{($10^{-5}$, $0$, $-5$)};
    
        \addplot [color=mycolor4, mark = x]
            table[]{figures/data/air_v_iter_64_qam_1e-5_10000km_13dB_-5dB.tsv};
        \addlegendentry{($10^{-5}$, $13$, $-5$)};
            
        \addplot [color=mycolor3, mark = o]
            table[]{figures/data/air_v_iter_64_qam_1e-3_10000km_0dB_-5dB.tsv};
        \addlegendentry{($10^{-3}$, $0$,$-5$)};
            
        \addplot [color=mycolor4, mark = o]
            table[]{figures/data/air_v_iter_64_qam_1e-3_10000km_13dB_-5dB.tsv};
        \addlegendentry{($10^{-3}$, $13$, $-5$)};
    \end{axis}
\end{tikzpicture}%
    }
    \caption{GMI vs. number of iterations. The legend shows $\sigma_\theta^2$, SNR [dB], and $20\log_{10}\rho$ [dB].}
    \label{fig:air_v_iter}
\end{figure}

\figref{fig:air_v_iter} plots \gls{gmi} of \gls{ep} vs. the number of iterations. Two iterations suffice to capture most of the performance gain over the feedforward receiver (one iteration). 

\section{Conclusions}
We studied \gls{pnc} prior to \gls{cdc} for feedforward and iterative \gls{ep}-based receivers, and compared with \gls{cdc} followed by \gls{idr}. The \gls{gmi} without \gls{eepn} is almost independent of fiber length.

Future work may consider using transmitter \gls{pn} and a dual-pilot tone to differentiate the transmitter and local oscillator \gls{pn}. One may also study more efficient \gls{pnc}, e.g., by using short and previously decoded blocks as side information to improve the feedforward receiver. Another interesting direction is the interaction with other distortions acting on the side information, e.g., polarization-mode dispersion in a dual-polarization system. 

\clearpage
\printbibliography

\end{document}